\documentclass[useAMs,a4paper]{mn2e}
\usepackage{savesym}
\usepackage{graphicx}
\expandafter\let\csname equation*\endcsname\relax
  \expandafter\let\csname endequation*\endcsname\relax 
\usepackage{subfig}
\usepackage{amsmath}
\usepackage{amssymb}
\usepackage{verbatim}
\usepackage[yyyymmdd,hhmmss]{datetime}
\usepackage{array}
\usepackage{times}
\usepackage[total={17.8cm,24.0cm},centering]{geometry} 
\usepackage{color}

\newcommand{\be}{\begin{equation}}
\newcommand{\beq}{\begin{equation}}
\newcommand{\ee}{\end{equation}}
\newcommand{\eeq}{\end{equation}}
\newcommand{\eea}{\end{eqnarray}}
\newcommand{\bea}{\begin{eqnarray}}

\newcommand{\dd}{\partial}

\newcommand\W {{W^r_{\ \phi}}}

\newcommand\WW {{\cal W}}

\title[Relativistic thin discs with finite ISCO stress]{Evolution of relativistic thin discs with a finite ISCO stress: \\
I. Stalled accretion}
\author  [Andrew Mummery, Steven A. Balbus]{Andrew Mummery\thanks{E-mail: andrew.mummery@physics.ox.ac.uk}, {Steven A. Balbus\thanks{E-mail:
steven.balbus@physics.ox.ac.uk}}
\\
Oxford Astrophysics, Denys Wilkinson Building, Keble Road, Oxford, OX1 3RH, United Kingdom}
\begin{document}

\date{}

\pagerange{\pageref{firstpage}--\pageref{lastpage}} \pubyear{2019}

\maketitle

\label{firstpage}

\begin{abstract} 
We present solutions to the relativistic thin disc evolutionary equation using an $\alpha$-model for the turbulent stress tensor.   Solutions with a finite stress at the innermost stable circular orbit (ISCO) give rise to bolometric light curves with a shallow power law time dependence, in good agreement with those observed in tidal disruption events.   A self-similar model based on electron scattering opacity, for example, yields a power law index of $-11/14$, as opposed to $-19/16$ for the case of zero ISCO stress.   These solutions correspond to an extended period of relaxation of the evolving disc which, like the light curves they produce, is not sustainable indefinitely.     Cumulative departures from the approximation of exact circular orbits cause the power law index to evolve slowly with time, leading eventually to the steeper fall-off associated with traditional zero ISCO stress models.   These modified solutions are discussed in detail in a companion paper.  

%The underlying equations governing thin-disc accretion discs are derived perturbatively in fluctuations in the velocity of the disc fluid. The standard assumption is that these perturbations are away from mean motion of precisely circular orbits. We demonstrate that in the presence of a finite ISCO stress -- something that has been observed in numerous numerical simulations -- this circular motion assumption leads to physically unsustainable results at large times.  A more sophisticated `quasi-circular' model of the mean fluid flow is required to accurately describe the evolution of relativistic thin discs. It is the goal of this paper to elucidate the need for this more sophisticated model, which is solved fully in a companion paper. We also present approximate solutions of the relativistic disc equation for an `alpha-disc' model of turbulence. 

\end{abstract}
\noindent
%Complied at \today\ \currenttime\ .

\section{Introduction}

The accretion of a debris disc formed when a wayward star has been disrupted by gravitational tides upon venturing too close to a supermassive black hole can produce X-ray flares from otherwise quiescent galactic centres.    These are {often referred to} as tidal disruption events, or TDEs.   In general, the late time luminosity emerging from such a TDE is expected to exhibit a light curve  $L(t)$ that varies as a power law in time $t$ with index $n$:  $L\sim t^n$.  This is of observational interest because the index $n$ can in principle be used as a model diagnostic.  A recent source compilation by Auchettl, Guillochon, \& Ramirez-Ruiz (2017; hereafter AGR) is revealing.  The late time power law indices of X-ray sources that AGR designate as ``confirmed'' centre around $n \simeq -0.75$.  This is a much more shallow fall-off than predicted by most theoretical models.   The original Rees (1988) fallback model, which assumed equal mass in equal energy intervals (and ballistic dynamics), leads to $n=-5/3$.    The post-disruption formation of a disc extends the duration of the emission somewhat; standard disc models (Cannizzo, Lee, \& Goodman 1990) typically yield an index of $n\simeq-1.2$.    The association of discs with TDEs has been recently strengthened, as evidence suggests that the late time evolution of many TDEs seems to be governed by accretion disc dynamics.  Van Velzen \textit{et al}. (2018) cleanly detected six TDE sources several years after their initial flares, and have found that the late time  FUV emission from these sources are well described by thermal emission from accreting discs. 

{ Recently, Balbus \& Mummery (2018, hereafter BM18) revisited the thin disc accretion models for TDEs.   These models have limitations, of course: 
there are many possible complications and additional components that may well be present in real discs, and prudence would suggest regarding this approach as a sort of baseline study. }   These authors found, somewhat surprisingly in view of previous results, that the observed shallow light curve power law values are in fact in good agreement with the BM18 evolving thin disc models.   While similar in some ways to the classic models put forward by Cannizzo {\em et al.} (1990), the new models differ in two important respects: the first is the inclusion of Kerr geometry in the disc evolution equation (Eardley \& Lightman 1975, Balbus 2017); the second is the imposed boundary condition of a finite stress at the innermost stable circular orbit (ISCO).    That these two ``refinements'' cause order unity changes seems counterintuitive:  the luminosity is dominated by the Keplerian region of the disc, so why should small-scale Kerr geometry and an altered inner boundary condition change the late time large scale emission in such a significant manner?     The point is that conceptually separating the inner regions and outer regions in this way is misleading: it is the requirement of a smooth joining of the inner relativistic disc onto the outer Keplerian region of the disc at larger radii that determines the required combination of outer disc modes that must be present over the extended course of evolution.   Modes that normally become singular at very small radius, and are therefore ruled out in a purely Newtonian treatment, must be retained in a relativistic treatment, which evinces no ISCO singularity.     The presence -- and eventual dominance -- of these previously neglected modes has significant observational consequences at later times. 

In this paper we develop and extend these findings in two important ways.   The first is to model the anomalous stress present in thin discs in a more general, and one might argue more physical, manner.   BM18 modelled the turbulent stress as a specified function of radius $r$.   In this paper, we solve the equivalent problem using an $\alpha$-disc formalism for the stress.   Essentially all analytic and semi-analytic treatments of accretion discs rely on the use of a so-called ``anomalous stress tensor'' to account for the enhanced angular momentum transport and thermal energy dissipation associated with the accretion process.   First invoked by Shakura \& Sunyaev (1973), a measured turbulent stress tensor may now be extracted from numerical simulations of discs, as a consequence of the magnetorotational instability (or MRI; Balbus \& Hawley 1991).   But even with explicit numerical studies, turbulence remains at  best poorly understood.   There is no consensus fundamental theory that allows one, for example, to express a stress tensor in terms of background mean disc properties.   There remains an ongoing need for the use of $\alpha$-based phenomenological modelling. 

The second and more far-reaching goal of this paper, as well as its companion (Mummery \& Balbus 2019; hereafter paper II), is to deepen our understanding of the mean fluid flow in the inner regions of relativistic discs by exploring the limits of the model.  We demonstrate that when a finite ISCO stress is present, which is natural when Maxwell stresses dominate the transport,  standard time-dependent disc modelling leads to an extended phase in which the accretion is effectively stalled.    While there is observational evidence for this in several TDE light curves, it cannot be the ultimate resolution of the accretion process.  By re-examining the assumption that fluid elements move precisely on Kerr circular orbits, we set the stage for a more general approach, developed in paper II, that leads to a long-term resolution of the stalling problem.    

One must exercise care with an asymptotically-ordered theory in which the angular momentum gradient of the mean motion of the fluid elements, a quantity which would pass through zero at the ISCO for pure circular motion, enters the governing evolution equation as a divisor:  any small deviation from a true circular orbit will impart a qualitative change in the disc evolution, first near the ISCO, but then propagating outward.   This interesting anomaly is not apparent in either vanishing stress or steady-state modelling.  It is only in time-dependent finite ISCO stress modelling that the consequences of a strictly circular orbit assumption become apparent.   We emphasise that the stalling behaviour is not in itself unphysical;  it emerges as a prolonged intermediate phase of disc evolution.     Indeed, not only is stalling supported by observations,  something very much like it may have already be seen in numerical disc simulations which have been run for a sufficiently long time (S\c{a}dowski \textit{et al}. 2015). 
 
The layout of this paper is as follows.   In \S 2, we describe the formal mathematical problem, which is to find the time-dependent behaviour of an initially localised ring evolving in accord with the relativistic thin disc evolutionary equation.    In \S 3, we present direct numerical integration of the idealised ``zeroth order solution'':  a disc evolving with turbulent stress described by an $\alpha$-disc model, which is either finite or vanishing at the ISCO, under the assumption that the mean azimuthal flow is precisely circular orbits.  The properties of the zeroth order solutions are discussed in \S 4, where it is demonstrated that a more careful model of the mean fluid flow is required to avoid behaviour that, while novel and not without numerical and observational support, is unsustainable in the long term.   Paper II discusses a possible resolution of this problem.   Finally, our conclusions are presented in \S 5.   Many of the more technical mathematical details pertinent to the analytic discussions are relegated to a set of four appendices.

\section{Analysis}\label{analysis}

\subsection{Review of governing equation and notation}
We seek the evolution of the azimuthally-averaged, height-integrated disc surface density $\Sigma (r, t)$, using standard cylindrical Boyer-Lindquist coordinates for a Kerr disc: $r$ (radial), $\phi$ (azimuthal), $z$ (vertical) and $t$ (time).   {For simplicity, the disc is assumed to lie in the $z=0$ symmetry plane.}   The contravariant four velocity of the disc fluid is denoted $U^\mu$; its covariant counterpart is $U_\mu$.  The specific angular momentum corresponds to $U_\phi$, a covariant quantity.     We assume that there is an anomalous stress tensor present, $\W$, due to low level disk turbulence.   This is a measure of the correlation between the fluctuations in $U^r$ and $U_\phi$ (Balbus 2017), and could also include correlated magnetic fields.  As its notation suggests, $\W$ is a mixed tensor of rank two.    It is convenient to introduce the quantity $Y$,
\beq\label{yy}
Y\equiv \sqrt{g}\Sigma \W =  r \Sigma \W,
\eeq
where $g>0$ is the absolute value of the determinant of the (mid-plane) Kerr metric tensor $g_{\mu\nu}$. The Kerr metric  describes the spacetime external to a black hole of mass $M$ and angular momentum $J$.  For our choice of coordinates, $\sqrt{g}=r$.   The ISCO radius is denoted as $r_I$, and $x\equiv r-r_I$.  Other notation is standard: {\em the speed of light is set to unity throughout,} the gravitational radius is  $r_g = GM$, and the black hole spin parameter $a = J/M$.

Under these assumptions, the governing equation for the evolution of the disc is quite generally given by (Eardley \& Lightmann 1974; Balbus 2017):
\beq\label{fund}
{\dd \Sigma\over \dd t} =  {1\over rU^0}{\dd\ \over \dd r}\left({1\over U'_\phi} \left[   {\dd Y \over \dd r}- U_\phi U^\phi (\ln\Omega)' Y \right]\right),
\eeq
where the primed notation $'$ denotes an ordinary derivative with respect to $r$, and $\Omega=d\phi/dt$.    A direct rendering (a sort of ``relativistic upgrade'') from the original Newtonian equation (Pringle 1981; Balbus \& Papaloizou 1999) would end with the first term on the right.   The second term is an additional $v^2/c^2$ relativistic correction, which becomes significant near the ISCO, stemming from the radiated photon angular momentum. 

\subsection {Compact formulation and general inner boundary considerations}

In this paper, we work entirely with equation (\ref{fund}).   Following Balbus (2017), we define $Q$ by
\beq
{dQ\over dr} = - U_\phi U^\phi (\ln\Omega)'.
\eeq
It may be quite generally shown that $e^{-Q} = U^0$ for particles undergoing circular motion in the equatorial plane of any azimuthally symmetric spacetime, which clearly includes {\bf the} Kerr geometry.  (See Appendix A3.)   With 
\beq
\zeta =Ye^Q = Y/U^0,
\eeq
equation (\ref{fund}) becomes 
\beq\label{27q}
{\dd\Sigma\over \dd t} =  {1 \over rU^0}{\dd\ \over \dd r} \left({U^0 \over U'_\phi}    {\dd\zeta \over \dd r}\right).
\eeq
In general,  the turbulent stress $\W$ may depend on the local surface density $\Sigma$, in which case the most general form of the evolution equation is
\beq\label{generalequation}
{\dd\zeta\over \dd t} = \left(\W + \Sigma {\dd\W\over \dd \Sigma} \right) {1 \over (U^0)^2}{\dd\ \over \dd r} \left({U^0 \over U'_\phi}    {\dd\zeta \over \dd r}\right).
\eeq

Some care is needed when specifying what exactly is meant by the 4-velocities and 4-momenta in equation (\ref{generalequation}).   Balbus (2017) splits the disc velocity $u^\mu$ into a mean component ($U^\mu$) and vanishing-mean fluctuating component ($\delta U^\mu$):
\beq
u^\mu = U^\mu + \delta U ^\mu ,
\eeq
with asymptotic scalings 
\beq\label{scalings}
\delta U_\phi \ll U_\phi, ~~~ U^r \ll \delta U^r \sim \delta U_\phi / r \ll rU^\phi .
\eeq
This is the key physical premise upon which the disc evolution equation is derived:  to zeroth order the fluid rotates about the origin, with fluctuations at first order due to the low-level disc turbulence.  Only at second order (in the fluctuation amplitude) does the fluid radial drift velocity emerge.

%These asymptotic scalings are the \textit{only} conditions placed on the 4-velocities of the disc fluid.   While it is standard to assume that the mean 4-vectors $U^0$ and $U_\phi$ are those associated with precisely circular motion (e.g. Novikov and Thorne 1973),  we shall see that holding to this approximation at all times and location is actually unsustainable.    The manner in which this approximation breaks down is important for understanding both numerical simulations and TDE observations.    
 
Following the assumption that the mean azimuthal flow corresponds to circular orbits,  the (spatial) amplitude of the well-behaved (temporal) Laplace-mode solutions of equation (\ref{27q}) must take the form of the Airy function derivative ${\rm Ai}'(-x)$ in the neighbourhood of the ISCO $x=0$ (BM18).    This function, which is exponentially cut-off inside the ISCO ($x<0$), together with the discarded branch Bi$'(-x)$ (exponentially growing for $x<0$), {\em both} have vanishing first derivatives at the ISCO.   This is therefore a mathematical requirement that must be applied to any numerical simulation, not a formal boundary condition {\em per se}.   The modal boundary condition is in the choice of an exponential cut-off in $\Sigma$, as opposed to explosive growth, moving inward of the ISCO.   The non-linear problem is not  amenable to a simple modal analysis.
In formulating non-linear solutions to equation (\ref{27q}), the question of the vanishing spatial gradient at the ISCO needs to be addressed with some care. 

We shall assume that the turbulent ISCO stress satisfies 
\beq
\W + \Sigma {\dd\W\over \dd \Sigma} \neq 0 ,
\eeq
a condition that must be valid for physical $\W >0$ and stable disc evolution $\dd\W/\dd\Sigma >0$.  {  (Conditions under which this may not be true, e.g. in the presence of the viscous Lightman-Eardley instability (1974), are discussed in \S\ref{ELinstability}. )}
The governing equation may be written in the form
\beq\label{eq}
{\dd\zeta\over \dd t} = {\cal W} {\dd\ \over \dd r} \left({U^0 \over U'_\phi}    {\dd\zeta \over \dd r}\right) ,
\eeq
where we have defined the stress-like quantity 
\beq
\WW= \WW(r,\zeta) \equiv  {1 \over (U^0)^2} \left(\W + \Sigma {\dd\W\over \dd \Sigma} \right) \geq 0 .
\eeq
It is straightforward to show that equation (\ref{eq}) may be rearranged to give: 
\beq\label{BC}
U_\phi ' \left[ U_\phi ' \dot \zeta - \WW U^0 \zeta '' + \WW \left(U^0\right) ' \zeta' \right] + \WW U^0 U_\phi ''  \zeta ' = 0 .
\eeq
(Here, a prime denotes a radial $r$ derivative and a dot a coordinate time $t$ derivative.)
As $U_\phi'(r_I) = 0$, if $\WW(r_I) \neq 0$ this equality can be satisfied at the ISCO if and only if $\zeta '(r_I) = 0$.  Thus, under the usual circular orbit assumption, the requirement that the first derivative of $\zeta$ vanishes at the ISCO is a generic requirement for any finite ISCO stress.     In the case of vanishing stress, $\WW(r_I) = 0$, the appropriate boundary condition is $\zeta(r_I) = 0$.   We shall henceforth refer to the latter as the ``vanishing stress boundary condition.''

\subsection {$\W$ in the $\alpha$-disc formalism}

We begin with the well-posed problem of an evolving Kerr disc under the assumption that the turbulent stress is given by the standard $\alpha$-disc prescription, and that the mean azimuthal flow corresponds to precisely circular orbits exterior to the ISCO.   
%The key paper I finding of a  are recovered with this more physical modelling for the stress, in particular the shallow power law fall off of the light curve $L(t)$.    We shall see, however, that the assumption of precise mean circular motion is unsustainable for arbitrarily large times, and that a simple modification of this assumption resolves the problem. 
The anomalous stress $\W$ serves both to transport angular momentum outwards as well as to extract the free energy of the disc shear, which is then thermalised and radiated from the disc surfaces.   In the $\alpha$-disc prescription, both processes are assumed to be described locally, and $\W$ has an {\em ad hoc} functional dependence on both the surface density $\Sigma$ and radius $r$.   For a general power law opacity of the form
\beq \label{opacity}
\kappa \propto \rho^a \, T^{-b} ,
\eeq
standard scaling arguments from classical disc theory (see Appendix A4) lead to:
\beq \label{stress}
\W \propto \Sigma^{(4+2a)/(6+2b+a)} ~r^{(3+2b-2a)/(6+2b+a)} .
\eeq
Accordingly, we work with a turbulent stress of the form
\beq \label{STRESS}
\W = w \left(\frac{\Sigma}{\Sigma_0}\right)^{\eta} \left(\frac{r}{r_0}\right)^{\mu},
\eeq
where $\Sigma_0$ and $r_0$ are fiducial values of the surface density and radius.   
Note that $w$ carries the dimensions of $\W$: (length)$^3$ (time)$^{-2}$. 
For the power law dependence of equation (\ref{STRESS}), the evolution equation becomes
\beq\label{gov}
{\dd\zeta\over \dd t} =  (1+\eta){\W \over (U^0)^2}{\dd\ \over \dd r}\left({U^0 \over U'_\phi}   {\dd\zeta \over \dd r}\right).
\eeq
{ Clearly $\eta > -1$ is required for stable disc evolution.   }

\subsection {Summary of previous analytical work}
The late time behaviour of transient relativistic discs may be calculated by a normal mode Laplace decomposition (BM18).  Whether the light curve is steep (power law index larger than 1 in magnitude) or shallow (index less than 1) depends upon which of the two possible Keplerian Laplace amplitudes dominates the luminosity profile at late times for a given ISCO stress boundary condition.    In the Keplerian regions of the disc, the general Laplace mode solution (time dependence $e^{-st}$) will have the general form 
\beq
\zeta(r,s) = c_1(s) \, \zeta_1(r,s) + c_2(s) \,\zeta_2(r,s),
\eeq
where $\zeta_1$ and $\zeta_2$ are linearly independent spatial amplitudes (BM18).    The $\zeta_2$ function vanishes at the origin along with its gradient, while only the gradient of $\zeta_1$ vanishes at $r=0$.    (We ignore for the moment the fact that $r=0$ lies inside of the Keplerian disc region.)  We shall always write the outer Laplace mode solution in this $c_1$, $c_2$ form: the  ``$c_1$ solution'' is understood to have a vanishing first derivative at the origin, while the ``$c_2$ solution'' itself vanishes at the origin. 
The choice of these coefficients (up to an overall normalisation) is determined by smoothly joining the outer solution to an inner modal solution local to the ISCO.  This inner solution in turn depends upon the selected ISCO boundary conditions, in particular whether or not the local stress vanishes.  

While the detailed modal solutions depend on the precise specification of the turbulent stress (i.e., the parameters $\eta$ and $\mu$), the gross features of $\zeta_1$ or $\zeta_2$ themselves are only slightly changed  over the parameter range of interest to this study.    By way of example, the simplest outer Newtonian solution ($\eta = \mu = 0$) is (BM18):
\beq\label{gensol}
\zeta =  r^{1/4}\left[ c_1(s) J_{-1/3}\left(pr^{3/4}\right) +c_2(s) J_{1/3}\left(pr^{3/4}\right)\right] ,
\eeq
where $J_{\pm1/3}$ are standard Bessel functions of fractional order, 
\beq
p^2 \equiv {\frac{s\sqrt{GM}}{3w}} ,
\eeq 
and $c_1$ and $c_2$ are the $s$-dependent constants for a given Laplace mode. In the relativistic problem, both the positive and negative Bessel index solutions will generally be required for a smooth match onto the inner ISCO solution.  The density distribution at a specified time $t$ is given by a superposition over all modes with a weighting function $f(s)$:
 \beq
 \zeta(r,t) = \int^\infty_0 f(s)\, e^{-st}\, \zeta(r,s) ~\text{d}s .
 \eeq
A particularly useful identity in this context is the Green's function linear superposition of Bessel function modes (Gradhsteyn \& Ryzhik 2014): 
$$
\int^\infty_0 J_\nu(\sqrt{s}R)J_\nu(\sqrt{s}R_0)  e^{-st}\, ds = \qquad\qquad\qquad\qquad\ \ \ \ \ \ \ \ 
$$
\beq\label{G2}
\ \ \ \ \ \ \ \ \ \  \qquad {1\over t}\exp\left(-R^2-R_0^2\over 4t\right)I_\nu\left(RR_0\over 2t\right) .
\eeq
At late times, the gross temporal behaviour of a disc's light curve will depend upon whether the $c_1$ or $c_2$ mode dominates.    { For a late time, finite ISCO stress disc, the $c_1$ solution ($\nu<0$) dominates; the $c_2$ solution ($\nu>0$) dominates in a vanishing stress disc (BM18).}  {  If
the late time luminosity behaviour of TDEs arises from an evolving disc, this is likely to be telling us something important about the ISCO stress: shallow power laws ($|n|<1$) imply a finite stress. } 

The work discussed in BM18 assumed a power law in $r$ dependence for the turbulent stress.  %as well as precisely Kerr metric circular motion. 
This paper extends this into an $\alpha$-disc regime, which allows a more general turbulent stress parameterisation of the form (\ref{STRESS}).   

The key question, both mathematically and physically, is whether we can understand the $\alpha$-disc behaviour in a manner analogous to the $c_1/c_2$ distinction that applies to the linear formulation.  Unlike a linear diffusion equation,  a non-linear evolution equation is not amenable to a Green's function treatment.    Fortunately, a set of analytic self-similar solutions to the $\alpha$-disc problem {\em are} known.    These govern the behaviour of the outer disc at late times in a manner very similar to the simpler linear disc equation, and are presented below.

\subsection {Newtonian self-similar solutions}
\subsubsection {General properties}\label{solutions}
There are self-similar solutions of the non-linear {\em Newtonian} limit of the disc evolution equation if the turbulent stress is parameterised in the  power law form given in equation (\ref{STRESS}). 
The solutions are conveniently expressed in terms of a dimensionless radius $x$ and time $\tau$:
\beq\label{taudef}
x = r/r_0, \quad \tau=  w t/\left(2\sqrt{GMr_0^3}\right).
\eeq
 (For the present $r_0$ is fiducial, but will be defined as part of our initial condition in \S 3.1.)  The mathematical solutions then have the general form (Pringle 1991)
\begin{align}\label{SSdef}
\Sigma &= \Sigma_0 \, f(\xi) \, x^{-3/2} \, \tau^{-\chi}, \\
 f(\xi) &= \xi^C (1-k\xi^B)^{1/\eta}, \\
 \xi &= \sqrt{x} \, \tau^{-\lambda} .
\end{align}
{ The five parameterisation constants ($B, C, k, \chi, \lambda$) all depend upon $\eta$ and $\mu$, and in fact there are two different self-similar solutions of this form, analogous to our $c_1$ and $c_2$ solutions, that pertain to different inner boundary conditions.   }This is discussed below.
 
These self-similar solutions are {\em exact} solutions of the non-linear Newtonian evolution equation in their own right, but they are not the most general, and they may be expected to be a good description of an evolving physical disc only at later times.   In effect, they govern the behaviour of the disc once the memory of the initial conditions has been lost.  The singular behaviour of these solutions at early times ($t \rightarrow 0$) is therefore not a concern.  In what follows, for some purposes we shall treat the self-similar solutions as if they were valid over the entirety of the disc, e.g. in analytically  calculating the disc luminosity.   Here, an exact treatment, which we carry out numerically, results in relatively small changes of detail.   
Following BM18, we expect the late time luminosity to be dominated by the self-similar solution associated with the appropriate inner boundary condition, which is set by the behaviour of the turbulent stress at the ISCO.

\subsubsection{ISCO feedback:  $c_1$ solutions}\label{FS}

Analogous to the two outer modal solutions in the linear evolution problem (both of which are generally needed, as we have seen), there are two self-similar solutions for each of the Newtonian limits of the non-linear evolutionary equation (\ref{gov}).   One of these solutions vanishes smoothly at small $r$ and is associated with the stress-free ISCO solutions (Cannizzo et al. 1990).    The other self-similar solution, with vanishing inner radial derivative  (Pringle 1991), is generally discarded. This self-similar solution is associated (again, in a Newtonian setting) with a hard inner torque, and it has a radial velocity which falls to zero at the inner disc edge.   The disc is then completely stalled.

For an evolving relativistic disc, the finite stress boundary condition also produces an unusual time-varying radial velocity across the ISCO that is reminiscent of disc ``stalling''.   This interesting and surprising behaviour of the radial velocity flow, which is key to the late time shallow fall-off in $L(t)$, is an indication that a more sophisticated model is required to describe the long-term evolution of a finite stress disc.  While the behaviour presented by these zeroth order solutions is not in itself  unphysical, it is not sustainable indefinitely.  A discussion of the underlying approximations that break down in these zeroth order disc models, and leads to this interesting behaviour, is given in \S 4. 

{ The five parameters which describe the self-similar solution (eq. \ref{SSdef})  satisfying} the finite stress boundary condition are given explicitly by:
\begin{align}
%A &= 1/\eta, \\ 
B &= (5\eta + 3 - 2\mu)/(\eta + 1),\\
C &= (3\eta + 1 - 2\mu)/(\eta + 1), \\
k &= \eta/[ (4\eta + 3-2\mu)(5\eta+3-2\mu)], \\
\lambda &= 1/(4\eta + 3 - 2\mu), \\
\chi &= 1/(4\eta + 3 - 2\mu).
\end{align}
This is mathematically equivalent to the Pringle (1991) solutions,  rewritten in the notation of the current paper.   At late times, the luminosity follows a power law decline (see Appendix A2):
\beq\label{finiteL}
L(t) \sim t^{-(4\eta + 2 - 2\mu)/(4\eta+3-2\mu)} .
\eeq
In terms of the opacity indices $a$ and $b$, this is
\beq\label{finiteLop}
L(t) \sim t^{-({22 + 14a)/({28 + 15a + 2b})}} .
\eeq
For example,  an electron scattering opacity ($a=b=0$) leads to a power law index of $-11/14\approx-0.79$.  Similarly, a disc dominated by Kramers opacity ($a = 1,~ b = 7/2$) leads to a power law index of $-18/25 = -0.72$.   The index is not highly sensitive to the scattering physics, and is robustly less than unity.

\subsubsection{Standard accretion: $c_2$ solutions }\label{VS}
If the turbulent stress vanishes at the ISCO, then so must the local surface density and luminosity.  In an earlier study of disc models for the late time luminosity arising from tidal disruption events,  this class of self-similar solution was analysed by Cannizzo \textit{et al}. (1990), and has become the canonical ``viscous model'' for TDEs.    Their parameters may be summarised as 
\begin{align}
%A &= 1/\eta ,\\
B &= (4\eta + 3 - 2\mu)/(\eta + 1), \\
C &= (3\eta + 2 - 2\mu)/(\eta + 1),  \\
k &= \eta/[(4\eta + 3 - 2\mu)(5\eta + 3 - 2\mu)],  \\
\lambda &= 1/(5\eta + 3 - 2\mu),  \\
\chi &= 2/(5\eta + 3 - 2\mu).
\end{align}
At late times, the luminosity follows a power law decline (Appendix A2),
\beq\label{vanishingL}
L(t) \sim  t^{-(5\eta + 4 - 2\mu)/(5\eta + 3 - 2\mu)} ,
\eeq
or equivalently 
\beq \label{vanishingLop}
L(t)  \sim t^{-({38 + 18a + 4b})/({32 + 17a + 2b})} .
\eeq
For the fiducial cases of electron scattering or Kramers opacity, the indices are $-19/16\approx -1.19$ and $-5/4=-1.25$ respectively.

It is interesting to note that Cannizzo \textit{et al}.\ set the emitted luminosity proportional to the rate of change of disc mass $\dot M$, whereas we  have performed a direct calculation of the locally emitted energy.   There is an exact agreement between our two methods for the vanishing stress case. The case of a finite ISCO stress however, essentially a model with feedback, requires an explicit calculation of the luminosity.   

\subsection{Recovery of linear theory from non-linear theory}
The results of BM18 are recovered in the linear $\eta \rightarrow 0$ limit. In this limit some care is required with handling the self-similar function  
\beq
 f(\xi) = \xi^C (1-k\xi^B)^{1/\eta} .
\eeq 
 In both the finite and vanishing stress cases as $\eta \rightarrow 0$ the coefficient $k \rightarrow 0$. Simultaneously the exponent of $f(\xi)$ diverges, so that some additional attention is required.  This involves making use of the well-known definition of the exponential function
\beq
\lim_{l \rightarrow \infty}  \left( 1 + {X}/{l} \right)^l  = e^{X}.
\eeq  
Using this, we rewrite the function $f(\xi)$ in equation (\ref{SSdef}) in a form appropriate for the linear case 
\beq
 f(\xi) = \xi^C (1-k\xi^B)^{1/\eta} \rightarrow \xi^C \exp \left[ -{{\xi^{B}}\over{(3-2\mu)^2}} \right]. 
\eeq
Using the above result, equation (\ref{SSdef}) becomes:
\beq
\Sigma \propto \frac{r^{-(\mu+3/4)}}{\tau} \exp\left[-\frac{(r/r_0)^{2q}}{(3-2\mu)^2~\tau}\right] \frac{r^{\pm 1/4}}{\tau^{\pm 1/4q}} ,
\eeq
where the $+/-$ sign is taken for the $c_2$/$c_1$ solutions respectively, and $q \equiv (3 - 2\mu)/4$ as defined in BM18.  Apart from a trivial difference in definition of $\tau$, this is exactly the expression for the late time expansion of the Green's function solutions of the linear evolution equation in BM18.  Even though no information of the initial condition is explicitly present in these self-similar solutions, the exact late time behaviour of an initially narrow disc centred at a distance $r_0$ from the black hole is recovered.  

\section{Numerical solutions}

The purpose of this section is to reexamine the BM18 solutions in the case of a non-linear version of the relativistic diffusion equation by numerical means.   The key findings to be verified are that there is a bimodal response of the late time luminosity behaviour of accretion discs -- the luminosity falls off at a rate either shallower or steeper than $t^{-1}$, depending upon whether the turbulent stress vanishes at the ISCO.  We shall also see that the actual late luminosity decay indices are well-approximated by those of the appropriate outer solutions of the Newtonian regime disc equation.    

 { The evolution equation in its compact has been used for the numerical integrations (eq. \ref{gov}).   The solution was found using essentially the same techniques described in Appendix A1 of BM18.    For $\alpha$-disc models, account needs to be taken of the non-linear character of the equation, and a predictor-corrector implicit method therefore must be used, rather than a simple implicit method. }
\subsection {Numerical formulation}
Our canonical run, which is more generally representative, is the evolution of a Gaussian ring at $t=0$.  The initial condition is:
\beq\label{initialcond}
\Sigma(r,t=0) = \Sigma_0 \exp\left[-\frac{(r-r_0)^2}{d^2}\right] .
\eeq 
The central radius $r_0$ is set to 15$r_g$, roughly the tidal radius of a solar mass star around a $4 \times 10^6$ solar mass black hole.   With $r_0$ set, $\Sigma_0$ and $d$ are related by the initial disc mass.  The initial spread is $d = 3.165$ gravitational radii, and the amplitude $\Sigma_0$ has been chosen so that the initial debris mass was 1 solar mass.   While the short-term evolution depends upon the initial radius and dispersion of debris, the long-term evolution of the extended disc does not.  All solutions were carried out for a rapidly (prograde) rotating black hole, $a/r_g = 0.9$.

At the inner edge of the evolving disc, corresponding to the ISCO, a boundary condition must be enforced, the precise nature of which depends upon the behaviour of the ISCO turbulent stress.   If a vanishing stress condition is imposed, the surface density (or equivalently $\zeta$) must vanish.   Assuming Kerr metric circular orbits, it is the radial gradient of $\zeta$ that must vanish if the ISCO stress is non-zero.    Numerically,  the value of $\zeta$ at the innermost ISCO grid point is then set equal to its value at the adjacent external grid point.

%\subsection {ISCO boundary conditions}

\subsection {Stress parameterisation}
We consider four different parameterisations of the turbulent stress behaviour: electron scattering and Kramer's opacity, with vanishing and finite ISCO stress.   
In our ``standard model'' the disc is dominated by electron scattering opacity and gas pressure. In this case, the opacity is a constant $\kappa_0$, with $a = b = 0$ in equation (\ref{opacity}).  This implies
\beq\label{standardmodel}
\W = w \left(\frac{\Sigma}{\Sigma_0}\right)^{{2}/{3}} \left(\frac{r}{r_0}\right)^{{1}/{2}} .
\eeq
The Kramer's opacity model assumes free-free or bound-free scattering throughout the disc.  The functional dependence of the opacity on temperature and density is $a = 1,~ b = 7/2$ (equation \ref{opacity}). This implies 
\beq\label{Kramersl}
\W = w \left(\frac{\Sigma}{\Sigma_0}\right)^{{3}/{7}} \left(\frac{r}{r_0}\right)^{{4}/{7}} .
\eeq 
As both stress parameterisations depend on a positive power of the surface density, it is sufficient to set the surface density to zero at the ISCO to enforce a vanishing stress boundary condition.

\begin{figure}
  \includegraphics[width=.5\textwidth]{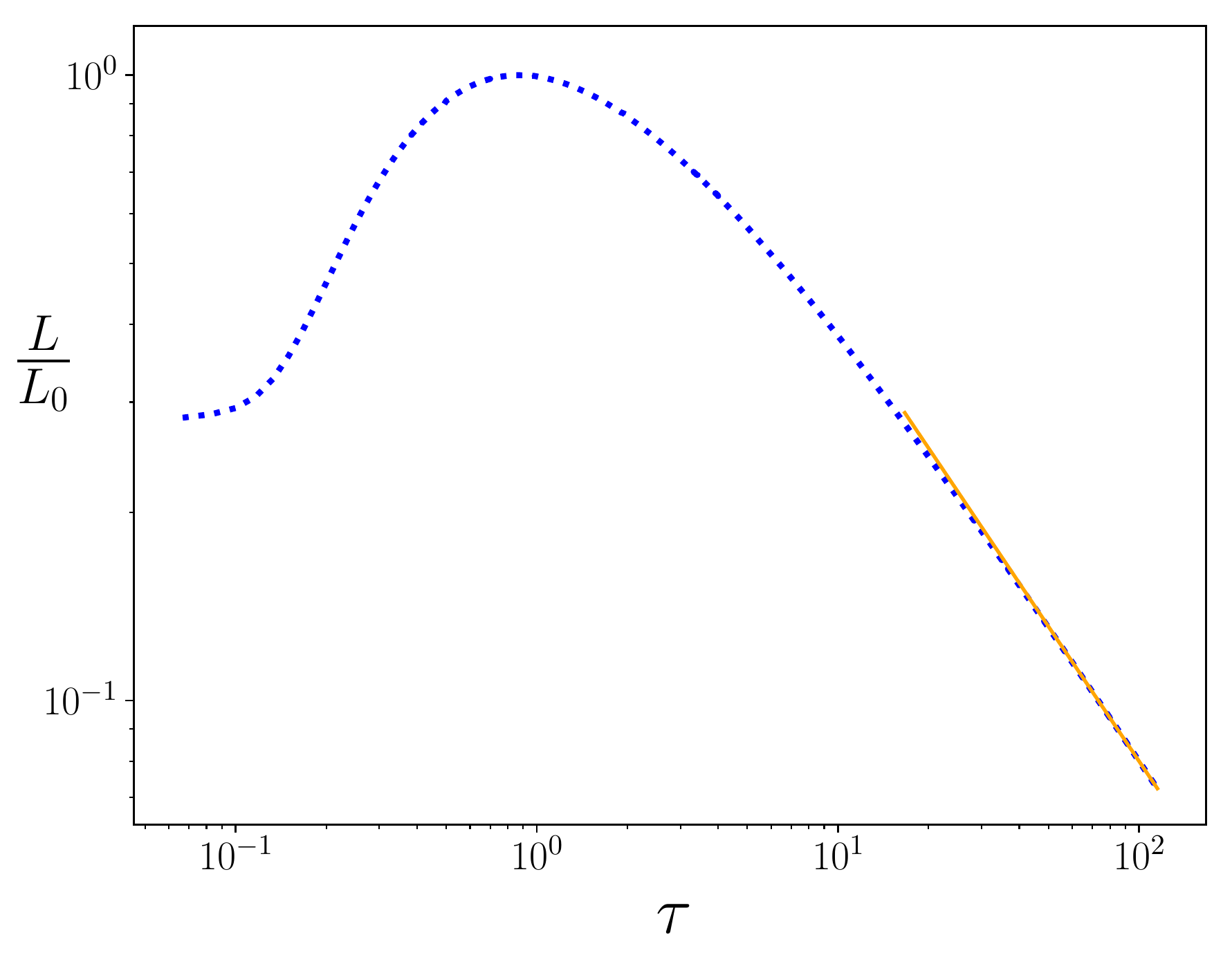} 
 \caption{The Luminosity $L(t)$ for the ``standard model'' (\ref{standardmodel}) with finite stress at the ISCO, based on solving the full Kerr equation (\ref{gov}). The profile is well fit by a (straight-line) power law at late times. The fitted late time behaviour is $L \sim \tau^{-0.72}$. The dimensionless viscous time (\ref{taudef}) is plotted on the x-axis. }
 \label{standardF}
\end{figure}

\begin{figure}
  \includegraphics[width=.5\textwidth]{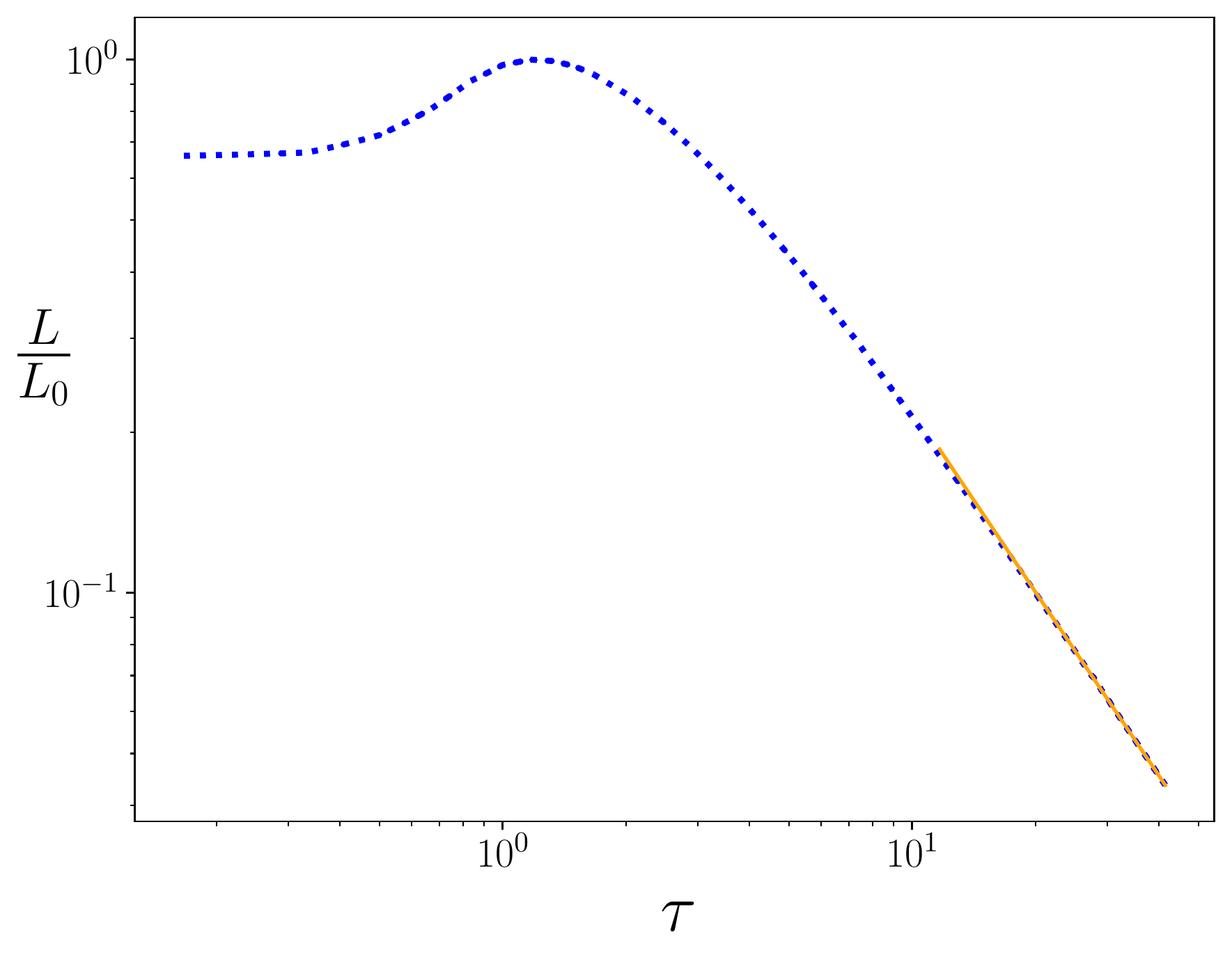} 
 \caption{The Luminosity $L(t)$ for the `standard model' (\ref{standardmodel}) with vanishing stress at the ISCO, based on solving the full Kerr equation (\ref{gov}). The profile is well fit by a (straight-line) power law at late times. The fitted late time behaviour is $L \sim \tau^{-1.14}$. The dimensionless viscous time (\ref{taudef}) is plotted on the x-axis.  }
 \label{standardV}
\end{figure}

\subsection{Summary and discussion}
{ We have solved numerically} for the time-dependent solutions of equation (\ref{gov}), the relativistic thin disc equation, under the assumption of Kerr metric circular orbits.  Two different inner boundary conditions were explored, that of vanishing stress at the ISCO and that of non-zero ISCO stress.  For the former $\zeta$ vanishes at the ISCO, whereas for the latter the radial gradient of $\zeta$ vanishes. 

Figs.\ (1) and (2) show example light curves for the case of electron-scattering opacity with the stress remaining finite (fig.\ [1]) or vanishing at the ISCO (fig. [2]).  
Table \ref{table1} presents a comparison between analytical predictions and numerical solutions for the  four different turbulent stress parameterisations considered in this section.  We see very good general agreement between prediction and simulations. An exact match would not be expected, as we have neglected any luminosity emitted from the inner strong field zone in the analytic modelling. Note, however, that regardless of the exact behaviour of the inner disc,  it must always join smoothly onto the self-similar solutions out at larger radius, and so it cannot have a significantly different time dependence. 

Recall that for the linear case treated in BM18, at late times the outer evolving disc is well approximated by one of two exponential self-similar solutions described earlier in this paper (eq.\ \ref{G2}).   It is therefore not surprising that in the non-linear theory, the late time behaviour of the disc is well modelled by the solutions presented in section \S \ref{FS} for discs with a finite ISCO stress, and by the solutions presented in section \S \ref{VS} for discs for a vanishing ISCO stress.

%QQQ

Once again, we find two very distinct late time behaviours for the integrated luminosity $L(t)$: the finite stress solutions fall off less rapidly than $t^{-1}$;  the vanishing stress solutions fall off more rapidly than $t^{-1}$.  This is insensitive to how the stress is modelled exterior to the ISCO.    Examination of equations (\ref{finiteL}) and (\ref{vanishingL}) shows that this bimodal behaviour in late time luminosity decay indices 
can be explained purely in terms of which set of self-similar solutions are present in the late time Newtonian disc regime.    This is a direct consequence of the properties of the turbulent stress at the ISCO.

 \begin{table}
\centering \begin{tabular}{|p{2cm} | c | c || c| c|} 
 \cline{2-5}
\multicolumn{1}{c|}{} &  \multicolumn{2}{p{2.2cm}||}{\centering  \hspace{.6cm} Finite \newline \textcolor{white}{hhh} ISCO Stress}  &\multicolumn{2}{ p{2.2cm}| }{\centering  \hspace{.4cm} Vanishing \newline \textcolor{white}{h}  ISCO Stress} \\
 \cline{2-5}
 \multicolumn{1}{c|}{}& $n~$(num) & $n~$ (eq. \ref{finiteLop})  & $n~$(num) &  $n~$ (eq. \ref{vanishingLop})  \\ [0.5ex] 
  \cline{2-5}
\hline
 Standard  Model & $ {-0.72}$  & $-0.79$& $-1.14$ & $-1.19$  \\  
  \hline

 Kramers Opacity & $-0.66$ & $-0.72$ & $-1.25$ & $-1.25$\\  
 \hline

\end{tabular}
 \caption{ Comparison between late time numerically determined luminosity decay indices $n$ (num) and the analytical $n$ values from equations  (\ref{finiteLop}) and  (\ref{vanishingLop}).  $\W$  is given either by electron scattering ``Standard Model'' (eq.\ [\ref{standardmodel}]), or by Kramer's opacity (eq.\ [\ref{Kramersl}]). Numerical solutions were found for blackhole angular momentum $a/r_g =0.9$, corresponding to rapid rotation. }
 \label{table1}
 \end{table}

These numerical results show that the findings of BM18 remain valid for the more general $\alpha$ parameterisations of turbulent stress (eq.\ \ref{STRESS}).

\section{A closer examination of the finite-stress disc solutions}
In the sense that they give a very good approximate description of the disc behaviour at large times, the self-similar solutions presented in \S\ref{FS} are proper mathematical solutions  of the relativistic evolution equation for the problem posed --- the evolution of a disc where the stress is allowed to be non-zero at the ISCO, and Kerr circular motion is assumed. The question is, are these solutions physically sustainable in a real thin disc?   
By examining the radial fluid motion within the evolving finite stress discs we will demonstrate that these ``zeroth order'' solutions are not indefinitely sustainable, and that a more sophisticated orbital model must be developed.  We examine here the underlying model assumptions  in the presence of a finite ISCO stress, and suggest how a more accurate, quasi-circular orbit model could in principle resolve the difficulties.   This more sophisticated model is then solved fully in a companion paper. 

\subsection{Radial fluid motion in a finite stress disc}\label{problem}
The notion that magnetic stresses could exert sizeable torques on the inner regions of accretion discs has a long history  (Page \& Thorne 1974, Krolik 1999, Gammie 1999).   Numerical studies over the past twenty years have generally shown that the magnetic fields that drive the MRI, the process thought to drive angular momentum transfer in discs, lead to extended evolutionary phases with non-zero stresses at the ISCO (e.g. Noble, Krolik, Hawley 2010). 

In a disc with a non-zero ISCO stress, angular momentum is transported from the unstable disc region ($r < r_I$), back into the stable disc region ($r > r_I$).  By the time they reach the event horizon, fluid elements (in numerical simulations) typically have angular momenta of order $5$--$15\%$ less than what is required of a circular orbit at the ISCO (Noble, Krolik, Hawley 2010).  This liberated angular momentum slows the rotation of the inner edge of the stable disc region, and results in the presence of an angular momentum flux from the ISCO neighbourhood.  As time progresses, this effective torquing of the inner disc edge leads to the dominance of our $c_1$ solution. 

In this picture, a continuous flow of disc material over the ISCO is required to produce the inner disc stress.  It is theoretically possible for a disc with no net accretion to have a non-zero torque at the ISCO, e.g. by field lines attached to a spinning blackhole threading the inner disc edge.
While such torques are potentially quite interesting, in this work our focus is upon the internal turbulent stress self-consistently produced by the disc material itself.     

Is this type of finite ISCO stress model physically sustainable?    To answer this question we consider the motion of the fluid elements within the disc. The fluid outside of the ISCO is assumed to move to leading order on circular orbits, at first order to fluctuate therefrom, and only at second order to exhibit a radial `drift' velocity.  The latter is caused by the turbulent stress of the product of correlated first order fluctuations.   This determines the radial disc evolution. This velocity hierarchy is encapsulated in the scaling arguments of equation (\ref{scalings}).
%\beq\label{scalings}
%\delta U_\phi \ll U_\phi, ~~~ U^r \ll \delta U^r \sim \delta U_\phi / r \ll rU^\phi .
%\eeq

To calculate the radial drift velocity, we use equation (29) of Balbus (2017), which follows from a combination of mass and angular momentum conservation:
\beq\label{angmombal}
\Sigma U_\phi '  U^r + {{1}\over{r}} {{\partial}\over{\partial r}} \left(r \Sigma \W \right) -  \Sigma \W U_\phi U^\phi \left( \ln \Omega \right) ' = 0 .
\eeq
Defining $\zeta$ as in \S \ref{analysis}, after a simple manipulation we have
\beq\label{UR}
U^r = - \frac{\W}{U_\phi '}\frac{ \zeta '}{\zeta} .
\eeq
If we assume Kerr metric circular orbits, then the ISCO values of both $U_\phi'$ and $\zeta'$ vanish, leaving -- by l'Hopital's rule 
\beq\label{FSur}
U^r(r\rightarrow r_I) = - \frac{\W}{U_\phi ''}\frac{ \zeta ''}{\zeta} ,
\eeq
where we note that 
\beq
U_\phi '' (r_I) > 0 .
\eeq
This result ensures that in a finite ISCO stress disc, unlike a vanishing ISCO stress disc, the radial velocity is formally non-infinite at the ISCO. 

\begin{figure}
  \includegraphics[width=.5\textwidth]{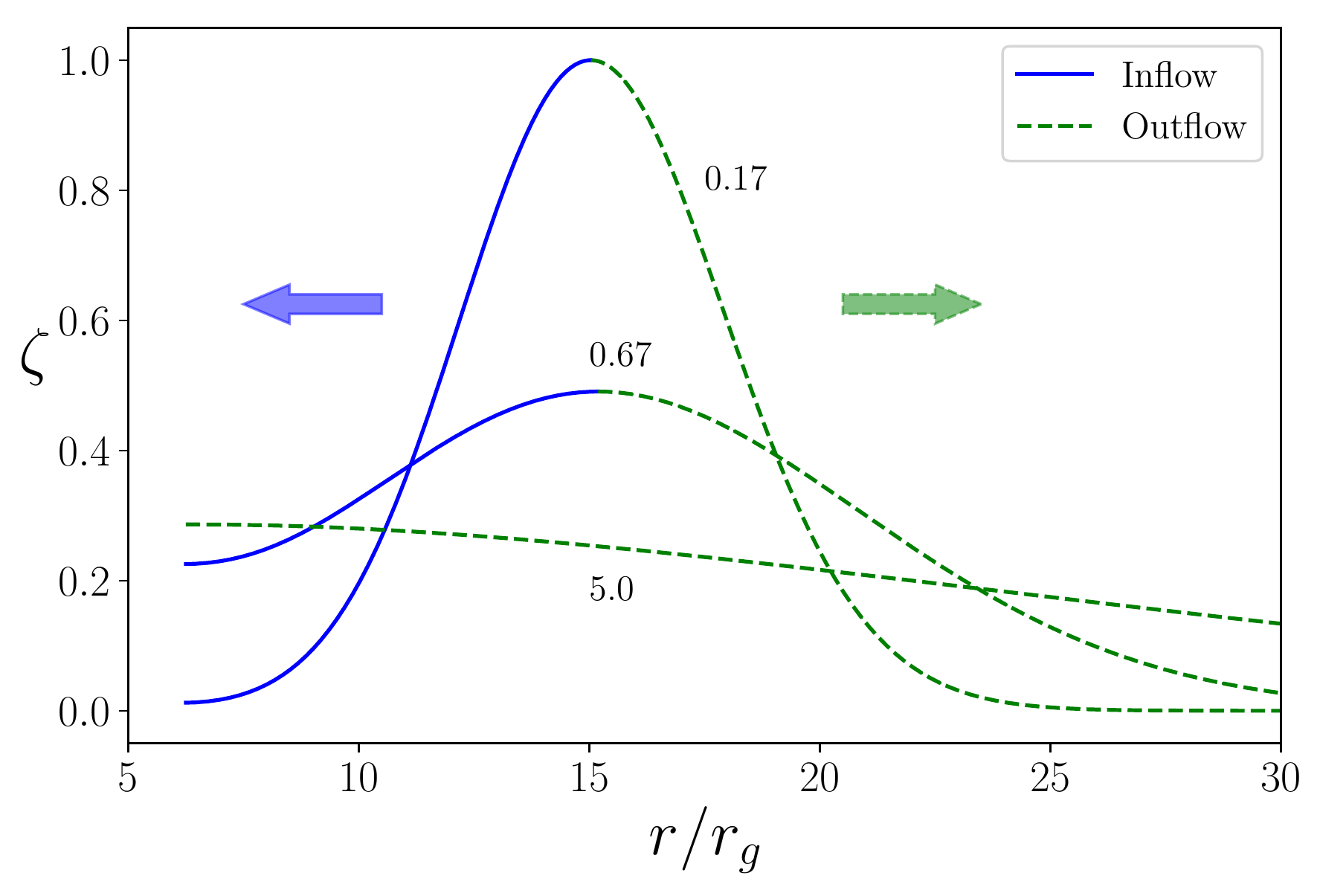} 
 \caption{Plots of the (normalised) dynamical variable $\zeta$ at three different times.    Each curve is labeled by its dimensionless time $\tau$.  The blue solid curves and green dashed curves represent regions in which the disc is inflowing and outflowing respectively. Eventually, $\zeta(r_I)$ becomes a global maxima, at which point the disc is outflowing at all external radii: the accretion stalls.}
 \label{fig3}
\end{figure}

Inspection of equation (\ref{UR}) demonstrates that the disc fluid flows in the direction of decreasing $\zeta$. Accretion is therefore cut off if $\zeta$ ever reaches a maximum at the ISCO, since this would produce a positive radial velocity at all points in the disc.  In the numerical finite ISCO stress solutions, this ISCO maximum is indeed seen (fig.\ \ref{fig3}). A systemic radial outflow at all points within the disc is an unsustainable result, as a real disc would be unable to maintain a non-zero ISCO stress indefinitely---not without continued accretion.  In a real disc, the lack of accreting gas flowing inward of the ISCO would almost certainly cause the ISCO stress to drop.  This would in turn lower $\zeta$, allowing accretion to pick up naturally once again. 

Clearly, a more careful model of the late time disc dynamics is required when the stress at the ISCO is non-zero. 

\subsection{Examining underlying assumptions}
The first term in this equation (\ref{angmombal}) describes an angular momentum deposition carried by the radial flow of the disc fluid.   The second term is the angular momentum flux divergence due to the various stresses present (turbulent, magnetic etc.), and the final term represents the angular momentum losses carried away by the emitted disc photons --- itself a by-product of the stresses acting on the disc. 

\begin{figure}
  \includegraphics[width=.5\textwidth]{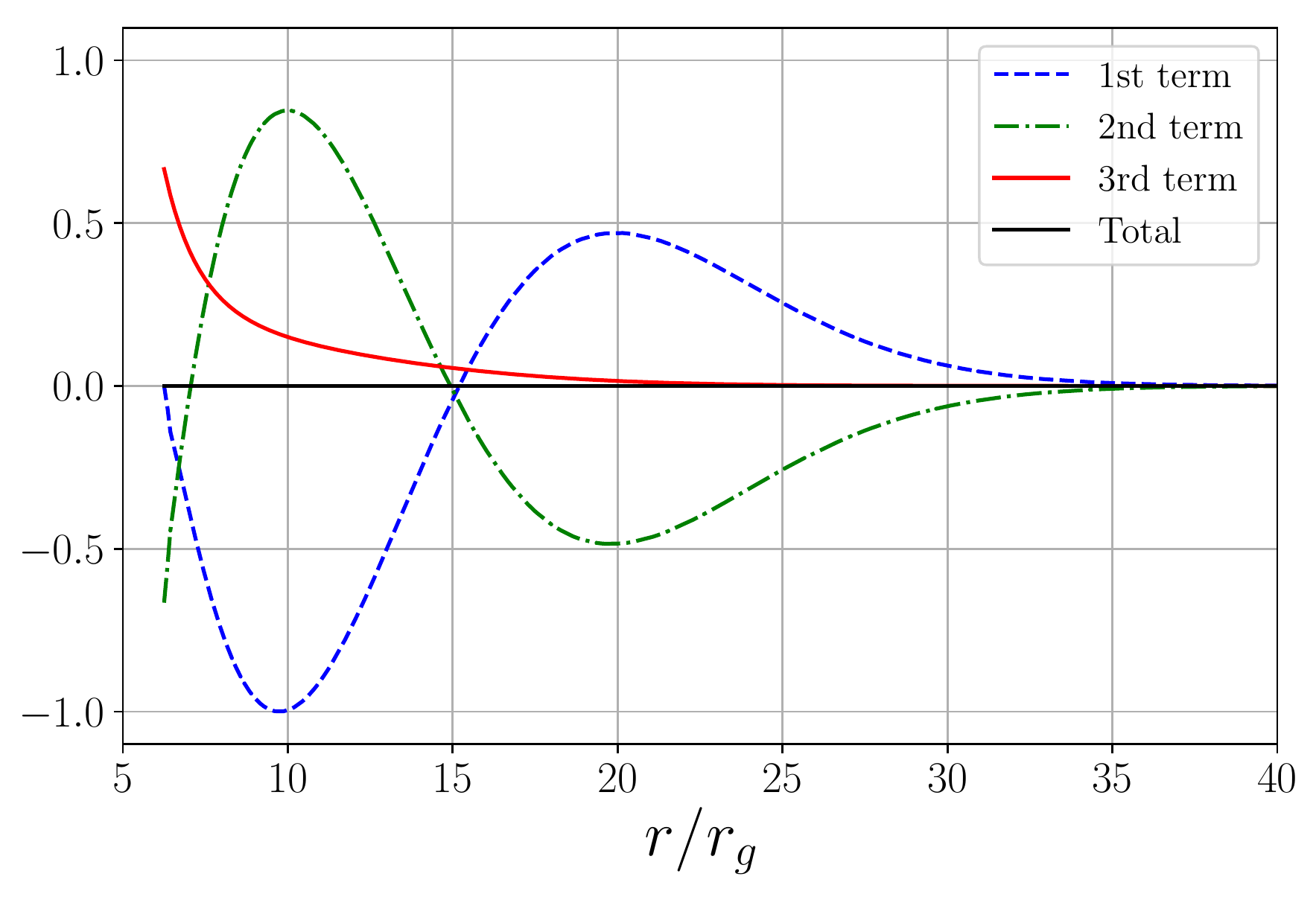} 
 \caption{The normalised angular momentum flux-divergences, stemming from the three terms in equation (\ref{angmombal}), at a particular time ($\tau = 0.67$) in the disc evolution.   Throughout the bulk of the disc, the angular momentum loss balance is predominantly between the radial flow (the first term) and the turbulent stresses (second term).  In the region $r \lesssim8 r_g$, however, the photon losses (third term) dominate over the radial flow stress.   This balance can be maintained as an extended phase, but not indefinitely.  }
 \label{fig4}
\end{figure}

It is clear from inspection of equation (\ref{angmombal}) that the Kerr circular orbit's vanishing ISCO angular momentum gradient locally conserves the radial flux of angular momentum arising from any mass flow.    If there is  a finite total  $\W$ stress present, then this model has a non-zero {\em positive} radial flux of angular momentum at a disc location where the angular momentum flux from the bulk disc flow is strictly conserved.   This outward  $\W$ flux is not conserved; its losses are a source for the photon angular momentum flux (fig.\ \ref{fig4}).   But in the model solution, the disc can only supply the $\W$ flux from the relatively small amount of material within the ISCO, because just outside the ISCO the disc velocity is outward: it is not accreting!  This cannot go in indefinitely---unless the outward $\W$ flux is somehow being fed externally.    The fact that the integrated total luminosity of our solutions formally diverges as $t\rightarrow \infty$ is a clue: something must be feeding this.     This could in principle be a magnetic couple to the black hole itself (or a vertical accretion process), but how would a turbulent/viscous model resolve this issue?   Our solutions must not be the true $t\rightarrow\infty$ disc solutions; they represent instead an extended phase of disc evolution in which the accretion is stalled.  While this may be very useful for understanding observations, it raises an important point of physics:  how does a real disc ``unstall'' itself?

\subsection{Modelling the angular momentum gradient in the inner disc}

The presence of a non-zero radial velocity means that the azimuthal motion is only approximately Kerr metric circular.    We have argued that the correction is second order in the fluctuation amplitude, but $U'_\phi$, the {\it derivative} of the angular momentum, appears as a singular denominator in the evolution equation.   Since strict adherence to the idealised model leads to long term unsustainable behaviour, it is of interest to consider 
a generic change to the angular momentum gradient of the form
\beq\label{newcirc}
U_\phi ' = \left(V^{c}_\phi \right)' + \epsilon (r) ,
\eeq
where now $U_\phi$ indicates the mean fluid angular momentum of equation (\ref{eq}), and $V^{c}_\phi$ is the idealised angular momentum of a Kerr circular orbit. The function $\epsilon(r)$ is left somewhat arbitrary, but it must satisfy two simple properties: (i) it reaches a maximum at the ISCO, and (ii) it vanishes sharply at large radius.  The idea is that the assumption of purely circular orbits is a good approximation at large radii, but become progressively worse as the ISCO is approached. 

The development of this theory is the topic of the companion paper.   We note here, however, that a modification of the angular momentum gradient in the inner disc regions would also affect the steady-state modelling of relativistic thin discs. The steady state solution of equation (\ref{eq}) leads to a disc temperature profile 
\beq
\sigma T^4 = - \frac{U^0U^\phi \ln\Omega '}{2r} \left[ F_1 \int\limits_{r_I}^{r} \frac{U_\phi '}{U^0} \,\text{d}r + F_2\right] ,
\eeq
where $F_1$ and $F_2$ are constants related to the mass and angular momentum flux throughout the disc respectively.  A substantial modification of $U_\phi'$ in the inner -- and therefore hottest -- disc regions would lead to changes in the predicted observational properties of the disc.

It is in time-dependent modelling where the effects of the modified angular momentum gradient are most apparent.  In the companion paper we solve, both analytically and numerically, the thin disc evolution equation (\ref{eq}) with an exponential form for $\epsilon(r)$.   This appears to fully resolve the issues highlighted in \S \ref{problem}, allowing for both an extended stalled phase of accretion, but ultimately a long-term return to a more standard disc model (and a finite luminosity integral).   Moreover, these new solutions display a rich set of behaviours which are compatible with the observed variations in TDE light curves.  The new solutions can no longer be described by just one of the sets of self-similar solutions in \S\ref{FS}  and \S \ref{VS}, but require instead characteristics of both in order to understand their behaviour. 

\subsection{The Lightman-Eardley instability }\label{ELinstability}
{An important, and possibly consequential, simplification used in this work (in common with Cannizzo \textit{et al}. 1990) is the neglect of the Lightman-Eardley (1974) instability in the inner regions of the disc, where the dynamical pressure becomes radiation-dominated.  If the turbulent stress is proportional to the dominant radiation pressure, as opposed to the gas pressure, the resulting sensitive dependence on temperature leads to a formal instability in $\alpha$ disc models. Although local 3D disc simulations with radiative hydro do exhibit unstable behaviour (Jiang, Stone, \& Davis 2013), as do global 2D hydro simulations (Fragile \textit{et al}. 2018), observations of X-ray binaries are compatible with stable thermal discs (e.g. Done, Gierli\'nski, \& Kubota 2007).   To what extent this process manifests as a true physical instability in real discs is unclear at the present time. }

{The point of view adopted in this work is that the Lightman-Eardley instability, if indeed present in nature, will result in {\em some} reasonably well-defined stress tensor $W^r_{\ \phi}$, perhaps not of the explicit $\alpha$-form used here, but in any case it will not completely disrupt the inner disc.  The principal conclusions of our study do not appear to be sensitive to the precise functional form of $W^r_{\ \phi}$:  so  as long as there exists some self-consistent local stress tensor, neglect of the formal Lightman-Eardley instability in the first instance seems a reasonable strategy.}

\section{Conclusions}
The work in this paper has served a dual purpose. The first is to extend the findings of BM18 into the (non-linear) $\alpha$-disc regime.   We have demonstrated that under the normal assumptions of thin disc theory, relativistic discs display bimodal late time behaviour, depending upon the nature of the turbulent stress boundary condition at the ISCO.   The luminosity emergent from finite ISCO stress discs falls off with time shallower than $t^{-1}$, as opposed to steeper than $t^{-1}$ for vanishing ISCO stress discs.  This may be understood quantitatively by examination of the two self-similar solutions of the underlying equation in its Newtonian limit.    The more shallow fall-off from finite stress solutions appears to be a much better match to the observed light curves of confirmed TDEs (Auchettl \textit{et al}. 2017).  

{It is important to note that finite bandwidth observations, even if broad, could produce light curves with somewhat different behaviour than the  formal bolometric luminosity presented here.   In a forthcoming paper (Mummery \& Balbus 2019, in preparation) this dependence on observational bandpass is examined in detail.   It is possible to derive useful closed form analytical expressions for the time-dependent luminosity in both UV and X-ray bands.   The mathematical solutions presented here are well adapted to a detailed spectral analysis, with the result that different disc models, particularly with regard to the presence or absence of an ISCO stress, may be distinguished.}
 %and we will show that these results still have a observable dependence on the power law index $n$ derived here, meaning that the differing models can still be tested and distinguished. }

Finally, it is as well to remember that the finite ISCO stress solutions represent only an extended phase of disc evolution, which is then altered on sufficiently long timescales.    Indeed, an important  finding of this paper is that resolving the puzzle of the long-term behaviour of finite ISCO stress discs requires consideration of corrections to the profile of the angular momentum gradient.       We demonstrate in a companion work (paper II) that introducing these modifications to the inner disc dynamics has profound effects on the long-term global behaviour of thin disc evolution, which appear to be compatible with a range of observations.  

\section*{Acknowledgments} 
It is a pleasure to acknowledge useful conversations with M.\ Begelman, J.\ Krolik, B.\ Metzger, and W.\ Potter.   This work is partially supported by STFC grant ST/S000488/1.

\subsection*{Appendix A1: Numerical calculations of the disc luminosity}

The derivation of the evolution equation (\ref{gov}) is based upon the assumption of small perturbations from circular orbits, and so within the ISCO it will quickly break down. Physically, we expect the fluid elements to quickly in-spiral after crossing the ISCO.  This occurs on a timescale similar to, or only somewhat longer than, the free-fall time. This was demonstrated by Shafee \textit{et al.\ } (2008), who found nearly laminar flow in full GRMHD simulations interior to the ISCO.  During this phase, the fluid elements emit almost no radiation and so hardly contribute to the disc's spectrum (Penna \textit{et al.\ } 2010).   Accordingly, numerical integration of equation (\ref{gov}) is restricted to the region of spacetime exterior to the ISCO, which is both mathematically self-consistent and physically sensible.  
 
Once the evolution equation has been solved, the time dependent luminosity is straightforward to calculate.  The local flux $\mathcal{F}$ from each face of the disc annulus is given by (Balbus 2017) 
 \beq
2\mathcal{F} = - \Sigma U^0 W^r_\phi \Omega ' .
\label{flux}
\eeq
In full detail the luminosity calculation is complicated.   Here we seek only the gross late time behaviour, not a detailed spectrum, and for this a simple face-on disc model is sufficient.   We  retain the gravitational and kinematic redshift effects, which introduce the ratio of observed to emitted flux,  $\left(U^0\right)^{-2}$, but neglect the photon orbit (``ray tracing'') details.   The total observed luminosity is then proportional to the integral
\beq
L(t) \propto \int\limits_{r_I}^\infty  \left( g_{rr} g_{\phi\phi}\right)^{1/2}{\mathcal{F}\over (U^0)^2} \, \text{d}r.
%\sqrt{g_{rr}g_{\phi\phi}} \left(\frac{1}{U^0}\right)^2 \mathcal{F}~ \text{d}r .
\eeq
More explicitly, this is 
\beq
L(t) \propto \int\limits_{r_I}^\infty \left(\frac{{r^2 + a^2 + 2r_ga^2/ r}}{{1 - {2r_g}/{r} + {a^2}/{r^2}}}\right)^{1/2} \frac{\zeta(r,t)}{r^{\frac{7}{2}}\left(1 + a\sqrt{{r_g}/{r^3}}\right)^2}~\text{d}r .
\label{luminosity}
\eeq
The integral in equation (\ref{luminosity}) was carried out using a standard Simpson algorithm. 

\subsection*{Appendix A2: Time dependence of luminosity emergent from self-similar disc solutions}
The self-similar solutions of the non-linear Newtonian disc equation all have the same functional form (eq.\ \ref{SSdef}), and the time-dependence of the luminosity at large times can be found straightforwardly.   In the Newtonian limit, the local emergent flux has the form 
 \beq
2\mathcal{F} = - \Sigma W^r_\phi \Omega ' ,
\label{Newtonianflux}
\eeq
and the total disc luminosity is given by
\beq\label{newtonianluminosity}
L(t) = \int\limits_{r_I}^{r_\text{out}} 4\pi r \mathcal{F}  \, \text{d}r .
\eeq
Technically, as the self-similar solutions are valid only in the outer Newtonian disc region, they should be used only down to a lower limit $r_N$, which is close to $r_I$ and represents the inner edge of validity.   We shall assume, however, that the late time dependence of the luminosity is dominated by the bulk of the disc material, and is not significantly changed by the detailed emission between the ISCO and $r_N$.     (This may of course be checked explicitly by working entirely with the exact solutions, as we have done.)  
%this is some radius at which this solution is no longer valid --- be that because it is the ISCO  or when the strong field solution to the evolution equation dominates the emitted luminosity more generally. 
The upper integration limit $r_\text{out}$ corresponds to the location of the evolving outer edge of the self-similar disc, and is given by 
\beq
r_\text{out} = r_0 \tau^{2\lambda } k^{-2/B} .
\eeq 
We next substitute the the general forms of the surface density solution (\ref{SSdef}) and turbulent stress  (\ref{STRESS}) into equation (\ref{newtonianluminosity}).    This produces the following integral
\begin{multline} \label{inttt}
L(\tau) = \frac{L_0 \tau^{-\gamma}}{r_0}  \int\limits_{r_I}^{r_\text{out}} \left( \frac{r}{r_0}\right)^{\xi}  \left[1 - k\left(\frac{r}{r_0}\right)^{B/2}\tau^{-\lambda B}\right]^{\delta} {\text{d}r},
\end{multline}
where we have defined $\gamma, ~\xi$ and $\delta$ as
\begin{align}
\gamma &= (1+\eta)(C\lambda + \chi), \\
\xi &= (1+\eta)\frac{(C-3)}{2} + \mu -\frac{3}{2}, \\
\delta &= (1+\eta)/\eta  .
\end{align}
We remind the reader that the parameters $C$, $\lambda$ and $\chi$ which appear in this luminosity integral depend both upon the specifics of the turbulent stress parameterisation ($\eta$ and $\mu$ in equation [\ref{STRESS}]), and also upon whether the stress vanishes or remains finite at the ISCO.   (\S 2.5.2 and \S 2.5.3.)

The integral may be expressed in terms of a hypergeometric function of $z = k\left({r_I}/{r_0}\right)^{B/2}\tau^{-\lambda B}$, but it is unwieldy.   As we are interested only in the behaviour of the emitted luminosity at large times, the leading order behaviour is sufficient. For physically relevant solutions of the disc equation,  $\lambda$ and $B$ are both greater than zero (Pringle 1991).   This means that for a given radius, the function in square brackets 
\beq
F(r,\tau) = \left[1 - k\left(\frac{r}{r_0}\right)^{B/2}\tau^{-\lambda B}\right]^{\delta},
\eeq
approaches unity at large times.   The luminosity of the disc solutions are dominated by the hot inner, but still robustly Keplerian, regions.   For sufficiently large times,  $F(r,\tau) \simeq 1$ over the bulk of the domain.  We may therefore read off the time dependence of the emitted luminosity from the lead $\tau^{-\gamma}$ factor: 
\beq
L \sim \tau^{-\gamma} \sim \tau^{-(1+\eta)(C\lambda+\chi)} .
\eeq
Substituting the appropriate values of $C$, $\lambda$ and $\chi$ from  \S \ref{FS}, the late time luminosity from  \textit{finite} ISCO stress discs is
\beq
L(t) \sim t^{-(4\eta + 2 - 2\mu)/(4\eta+3-2\mu)}.  
\eeq
By contrast, the vanishing stress solutions (\S \ref{VS}) follow 
\beq
L(t) \sim  t^{-(5\eta + 4 - 2\mu)/(5\eta + 3 - 2\mu)} .
\eeq
These behaviours are in very good agreement with exact numerical calculations.

\subsection*{Appendix A3: Proof that $e^{-Q} =U^0$}
Start with
\beq
{dQ\over dr}\equiv - U_\phi U^\phi (\ln\Omega)',
\eeq
where the primed ($'$) notation on the right denotes $d/dr$.  
With 
\beq
\Omega\equiv  {d\phi\over dt} = {U^\phi\over U^0},
%\ln\Omega ' = ( U^0 U^\phi {}' - U^\phi U^0 {}' )/( U^\phi U^0 ),
\eeq 
we  find
\beq\label{aa31}
{dQ\over dr} = - U_\phi (U^\phi {})' + U_\phi U^\phi (U^0 {})' / U^0 .
\eeq
For a particle undergoing circular motion in the equatorial plane, 
\beq\label{aa32}
U_\mu U^\mu = U_\phi U^\phi + U_0 U^0 = -1.
\eeq
Substituting for $U_\phi U^\phi $ from (\ref{aa32}) into (\ref{aa31}),
\beq\label{dqdr}
{dQ\over dr} = - \left(\ln U^0\right)' -\left[ U_\phi (U^\phi {})' + U_0 (U^0 {})' \right]. 
\eeq
To prove that the second grouping on the right vanishes for circular orbits, consider the covariant null 4-vector
\beq
X_\mu = g_{\alpha\beta} U^\alpha U^\beta_{; \mu}  = {1\over 2} (U^\alpha U_\alpha)_{; \mu} \equiv 0 .
\eeq
The middles equality follows as $g_{\alpha\beta}$  is symmetric and $g_{\alpha \beta ;\mu}= 0$.  (The semi-colon as usual denotes a covariant derivative.)    Expanding the covariant derivative in the first term on the right, we find
\beq
X_\mu = g_{\alpha\beta}U^\alpha \partial_\mu U^\beta + g_{\alpha\beta}\Gamma^\beta_{\sigma \mu} U^\alpha U^\sigma .
\eeq 
where $\Gamma^\beta_{\sigma\mu}$ is the usual affine connection.  
Writing the final term explicitly, with $\dd_\sigma\equiv \dd/\dd x^\sigma$:
\beq
g_{\alpha \beta}\Gamma^\beta_{\sigma \mu} U^\alpha U^\sigma = \frac{1}{2} \left( \partial_\sigma g_{\mu \alpha} + \partial_\mu g_{\sigma \alpha} - \partial_\alpha g_{\mu\sigma} \right)U^\sigma U^\alpha .
\eeq
The first and third terms in the above cancel by symmetry.  We therefore conclude
\beq
X_\mu = g_{\alpha\beta}U^\alpha \partial_\mu U^\beta + \frac{1}{2} \left(\partial_\mu g_{\alpha\beta} \right)U^\alpha U^\beta = 0 .
\eeq
The final term vanishes with the help of the geodesic equation for $U_\mu$, 
\beq
{dU_\mu\over d\tau} = \frac{1}{2} \left(\partial_\mu g_{\alpha \beta} \right)U^\alpha U^\beta,
%X_\mu = g_{ab}U^a \partial_\mu U^b + \dot U_\mu .
\eeq
which for $\mu=r$ must be zero for circular orbits.  Hence,
%The radial component of this vector is simply
%\beq
%X_r = g_{ab}U^a \partial_r U^b + \dot U_r = 0 ,
%\eeq
 %for a particle on a circular orbit $\dot r = 0$, which implies $ \dot U_r = 0$, and so
\beq
X_r = g_{\alpha\beta}U^\alpha \partial_r U^\beta = U_\phi (U^\phi {})' + U_0 (U^0 {})'  = 0 .
\eeq
Equation (\ref{dqdr}) then becomes
\beq
{dQ\over dr} = - {d \over dr} \left(\ln U^0\right) , 
\eeq
which completes the proof, since we are free to choose $Q\rightarrow 0$ as $U^0\rightarrow 1$.  (Note that the value of the proportionality constant is not actually needed in the governing equation.)

\subsection*{Appendix A4: Surface density and radial dependence of $\W$ in an $\alpha$-disc model}
We start with the statement that on extended evolutionary time scales, the energy locally extracted from the disc shear must balance the locally radiated energy:
\beq\label{Ebal}
\W \Sigma \Omega ' \propto \frac{T^4}{\tau} \propto \frac{T^4}{\kappa \Sigma} .
\eeq
Here $\kappa$ is the opacity, $\tau$ the optical depth, $\Sigma$ the surface density, $\W$ the turbulent stress, $\Omega$ the angular frequency, and $T$ the mid-plane temperature.   Note that a standard $\alpha$-disc model is defined through the phenomenological relationship (Shakura \& Sunyaev 1973):  
\beq\label{eq1b}
\widetilde W_{r\phi} \equiv \alpha c_s^2 .
\eeq

The original Shakura \& Sunyaev notation included a factor of the surface density $\Sigma$ on the right, but more importantly, $\widetilde W_{r\phi}$ should not, despite the presence and placement of the indices, be regarded as a true covariant tensor (hence the tilde notation).    Following the standard formalism of equation (\ref{eq1b}),  $\alpha$ is assumed to be constant throughout the disc, and $c_s$ is the local sound speed of the orbiting gas.   $\widetilde W_{r \phi}$ measures correlations between ``ordinary'' turbulent \textit{velocity} fluctuations, whereas $\W$ measures correlations between fluctuations in the fluid angular momentum and radial velocity, a true 4-velocity correlation.  The two quantities are related by 

\beq\label{eq1}
\widetilde W_{r\phi} = \W / r \propto c_s^2 \propto T .
\eeq
In the final proportionality on the right, we have used the gas pressure to compute the sound speed, an important assumption that may, as has been noted, break down when the total pressure is dominated by radiation.   (In fact, it is probably more accurate to use a magnetic pressure.)    In section \ref{ELinstability}, we discuss in more detail the role of a possible viscous instability that emerges from the use of radiation pressure in (\ref{eq1}).   Since our principal conclusions are insensitive to the precise functional form of $\W$, we follow the simplest path here, bearing in mind its possible limitations.  

The disc scale height $H$ is related to the disc sound speed by
\beq
c_s \propto H \Omega \propto T^{1/2}.
\eeq
We assume a disc opacity of the form
\beq
\kappa \propto \rho^a\,T^{-b} ,
\eeq
which may be used for a wide range of physically plausible models, including electron scattering $a=b=0$, or a Kramers opacity: $a = 1, b = 7/2$. The density and surface density of the disc are related through the disc scale height
\beq
\rho \propto \Sigma / H .
\eeq
Finally, assuming Keplerian rotation 
\beq\label{eq2}
\Omega \propto r^{-3/2} ,
\eeq 
these relationships suffice to define the parameter dependence of the turbulent stress. Substitution of equations (\ref{eq1} -- \ref{eq2}) into equation (\ref{Ebal}) leads to the result quoted in the paper
\beq
\W \propto \Sigma^{A} \, r^{B} .
\eeq
Where
\beq 
A = (4+2a)/(6+a+2b) ,
\eeq
and
\beq
B = (3+2b-2a)/(6+a+2b) .
\eeq
\label{lastpage}

\end{document}